\documentclass[superscriptaddress,twocolumn,showpacs,amsmath,amssymb]{revtex4}

\usepackage{graphicx}
\usepackage{dcolumn}
\usepackage{bm}
\begin{document}

\preprint{}

\title{Transition from subbarrier to deep subbarrier regimes in heavy-ion fusion reactions}

\author{Ei Shwe Zin Thein}
\affiliation{
Department of Physics, Mandalay University,
Myanmar}

\author{N.W. Lwin}
\affiliation{
Department of Physics, Mandalay University,
Myanmar}

\author{K. Hagino}
\affiliation{
Department of Physics, Tohoku University,
Sendai 980-8578, Japan}

\date{\today}

\begin{abstract}
We analyze the 
recent experimental data of heavy-ion fusion cross sections 
available up to deep subbarrier energies in order to discuss 
the threshold incident energy for a deep subbarrier fusion hindrance phenomenon. 
To this end, we employ 
a one-dimensional potential model with a Woods-Saxon internuclear potential. 
Fitting the experimental data in two different energy regions with 
different Woods-Saxon potentials, we define the threshold energy as an intersect of 
the two fusion excitation functions. 
We show that the threshold energies so extracted are 
in good agreement with the empirical systematics as well as with 
the values of the Krappe-Nix-Sierk (KNS) potential at the touching point. 
We also discuss the asymptotic energy shift of fusion cross sections 
with respect to the potential model calculations, and show that 
it decreases with decreasing energies in the deep subbarrier region 
although it takes a constant value 
at subbarrier energies. 
\end{abstract}

\maketitle

Heavy-ion fusion reactions at low incident energies 
are intimately related to 
the quantum tunneling phenomena of many-body systems. 
Because of a strong 
cancellation between the repulsive Coulomb interaction and an attractive short range 
nuclear interaction between the colliding nuclei, a potential barrier, referred to as a 
Coulomb barrier, is formed, which has to be surmounted in order for fusion to take place. 
In heavy-ion reactions, because of a strong absorption inside the Coulomb barrier, it
has been usually assumed that a compound nucleus is automatically formed once the
Coulomb barrier has been overcome. 
The simplest approach to heavy-ion fusion reactions based on this idea, that is, a one-dimensional 
potential model has been successful in reproducing experimental fusion cross sections 
at energies above the Coulomb barrier\cite{W73}. 
A one dimensional potential model fitted to reproduce fusion cross sections above 
the Coulomb barrier, however, have been found to 
underestimate fusion cross sections at lower energies. 
It has been well recognized by now that 
the sub-barrier fusion enhancement is caused by 
couplings of the relative motion between the colliding nuclei 
with other degrees of freedom, such as collective vibrational and rotational 
motions in the colliding nuclei\cite{DHRS98,BT98}. 

The behavior of fusion cross sections at extremely low 
energies is a critical issue for estimating reaction rates 
of astrophysical interests. 
One of the currents interests in heavy-ion fusion reactions is a steep fall-off 
phenomenon of fusion cross sections at deep subbarrier energies. 
Recently, fusion cross sections for several colliding systems have 
been measured down to extremely low cross sections, 
up to several nb\cite{bt8,bt5,Stefanini08,Montagnoli12,ANU07}.
These experimental data have shown that 
fusion cross sections fall off 
much more steeply 
at deep subbarrier energies as decreasing energies, compared to 
the expectation from the energy dependence of cross sections at subbarrier energies. 
Although a few theoretical models 
have been proposed\cite{IHI09,ME06}, the origin for the deep subbarrier fusion 
hindrance has not yet been fully understood. 

In Refs.\cite{bt8,bt5,bt9}, the deep subbarrier fusion hindrance has been 
analyzed using the astrophysical S-factor. 
It has been claimed \cite{bt8,bt5,bt9} 
that the deep subbarrier fusion hindrance sets in at the energy 
at which the astrophysical S-factor takes the maximum. 
The authors of Refs.\cite{bt8,bt5,bt9} even parametrized the threshold energy as 
\begin{equation}
E_s= 0.356\left(Z_1Z_2\sqrt{\frac{A_1A_2}{A_1+A_2}}\right)^{2/3}~~~({\rm MeV}).
\label{eq:E_s}
\end{equation}
Notice that the S-factor representation provides a useful tool only when the penetration 
of the Coulomb 
repulsive potential is a dominant contribution, such as in fusion reactions of light systems 
at low energies.  
In fact, the relation between the threshold for the deep subbarrier hindrance 
and the maximum of the S-factor is not clear physically, and thus it is not trivial 
how to justify theoretically the identification of the threshold energy with the astrophysical S-factor. 
Nevertheless, it has turned out that the threshold energy so obtained closely follows 
the values of phenomenological internucleus potentials, such as the 
Krappe-Nix-Sierk (KNS)\cite{KNS}, 
the Bass\cite{Bass}, the proximity\cite{proximity}, and 
the Aky\"uz-Winther\cite{AW} potentials, at the touching configuration \cite{bt1}. 
This clearly implies that the dynamics which takes place after the colliding nuclei touch with each other 
is responsible for the deep subbarrier fusion hindrance, making at the same time the 
astrophysical S-factor decrease as the incident energy decreases. 

In this paper, 
we investigate the threshold energy for deep subbarrier fusion hindrance using 
an alternative method, which 
is physically more transparent than the definition with the maximum of 
S-factor. 
That is, we determine the threshold energies by fitting the experimental fusion 
cross sections in subbarrier and deep subbarrier energy regions separately 
using single-channel barrier penetration model calculations, 
and compare them with the systematics given by 
Eq. (\ref{eq:E_s}) as well as with the touching energy evaluated with the KNS potential. 
We also discuss the energy dependence of fusion cross sections at deep subbarrier energies 
in terms of an asymptotic energy shift proposed by Aguiar {\it et al.}\cite{bt3}. 

\begin{figure}[tbhp]
\begin{center}\leavevmode
\includegraphics[width=1\linewidth, clip]{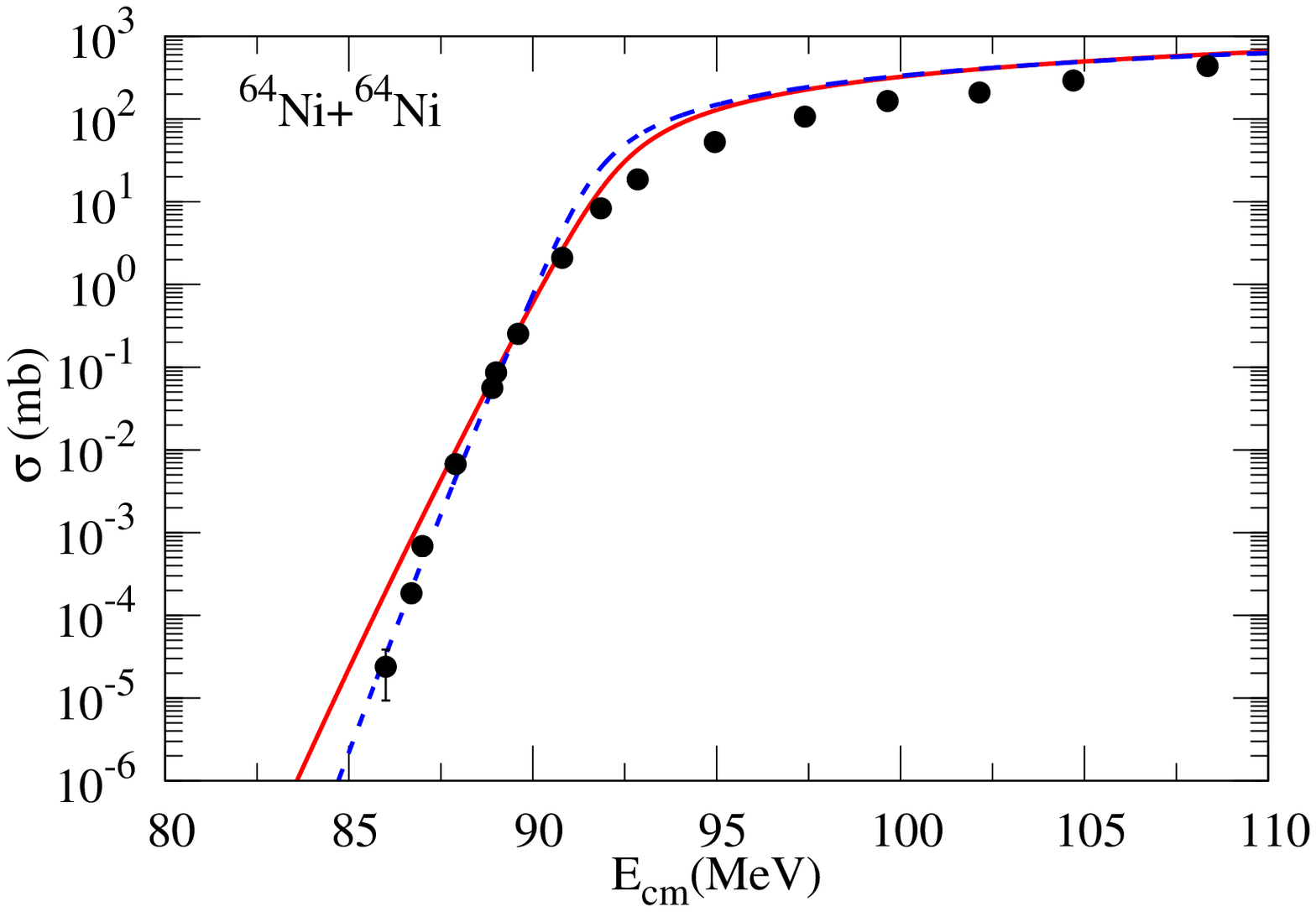}
\includegraphics[width=1\linewidth, clip]{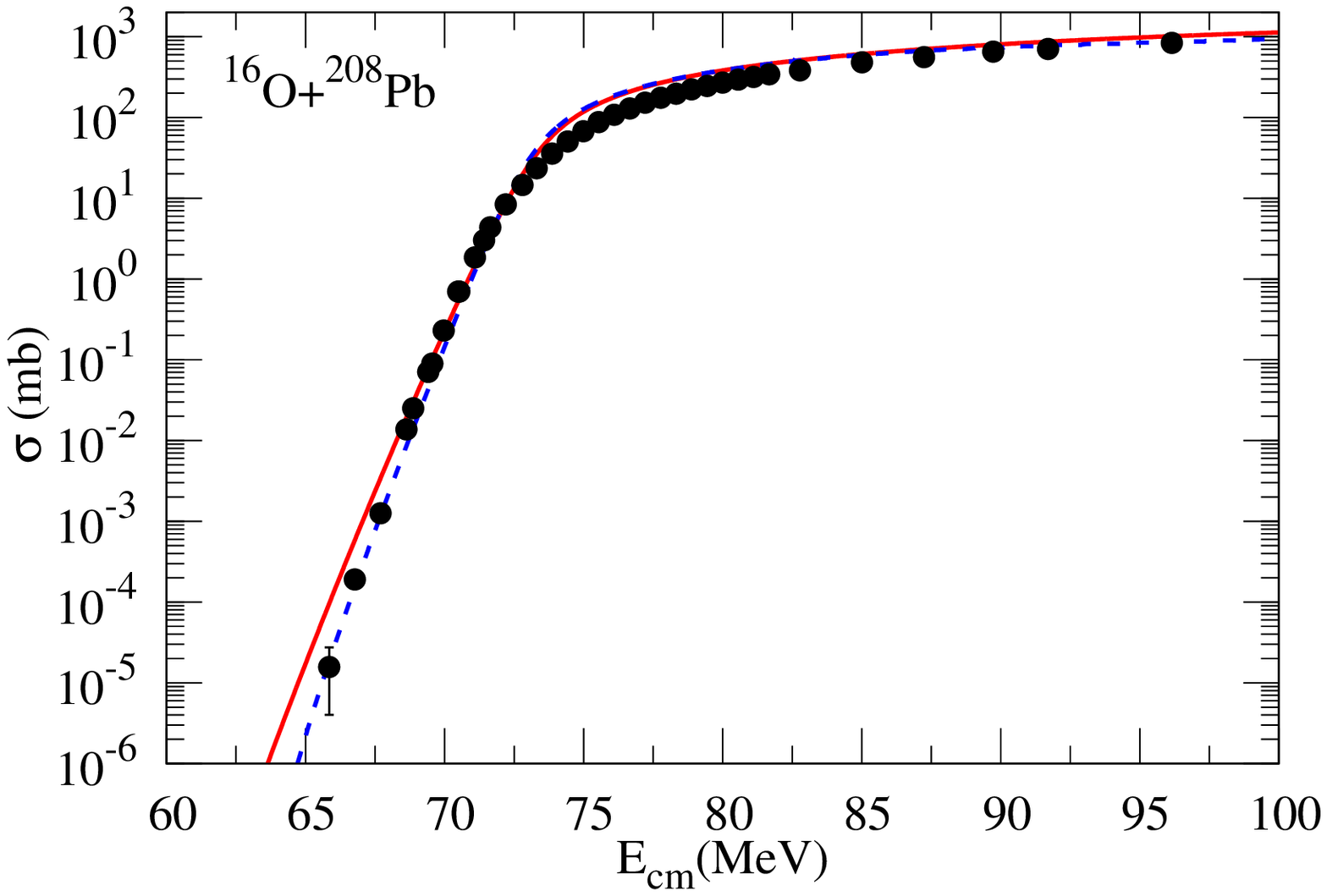}
\caption{(Color online) 
Fusion excitation functions for the 
$^{64}$Ni+$^{64}$Ni (the upper panel) and $^{16}$O+$^{208}$Pb (the lower panel). 
The solid and the dashed lines are results of single-channel 
potential model calculations which fit the experimental data in the subbarrier and the 
deep subbarier energy regions, respectively. The experimental data are taken from 
Refs. \cite{bt8,ANU07}. }
\label{fig:1}
\end{center}
\end{figure}

In order to illustrate our procedure, the upper and the lower panels of 
Fig. \ref{fig:1} show fusion cross sections for 
$^{64}$Ni+$^{64}$Ni and $^{16}$O+$^{208}$Pb systems, respectively. 
We first define the subbarrier energy region as the one in which fusion 
cross sections take between $10^{-2}$ mb and $10^0$ mb.
We fit the experimental data in this energy region with a potential model 
with a Woods-Saxon potential treating the three parameters of the potential, that is, 
the depth $V_0$, the radius $R_0$, and the surface diffuseness $a$, as adjustable 
parameters. To this end, we numerically solve the Schr\"odinger 
equation without resorting to the parabolic approximation\cite{bt7}. 
The fusion cross sections calculated in this way 
are shown by the solid lines in the figure. 
Of course, these calculations do not account for the fusion cross sections at 
higher energies as the channel coupling effects are completely ignored. 
However, it is sufficient for our purpose, as we are interested only in the 
energy dependence of fusion cross sections at subbarrier energies, 
that is, the slope of fusion excitation 
functions. These calculations do not reproduce the experimental data at lower 
energies, either. In order to obtain a better fit in the lower energy region, the 
surface diffuseness parameter has to be increased, as has 
been noticed in Refs. \cite{ANU07,bt7}. 
We then define the deep subbarrier region as the one in which fusion cross sections 
take below $10^{-3}$ mb. The dashed lines in the figure show the fusion 
cross sections obtained by fitting to the experimental data in this energy region. 
See Table I for the actual values of the surface diffuseness parameter. 
From the two curves, 
we finally define the threshold energy for deep subbarrier fusion 
hindrance as the 
energy at which the two fusion excitation functions intersect with each other. 

\begin{figure}[tbhp]
\begin{center}\leavevmode
\includegraphics[width=1\linewidth, clip]{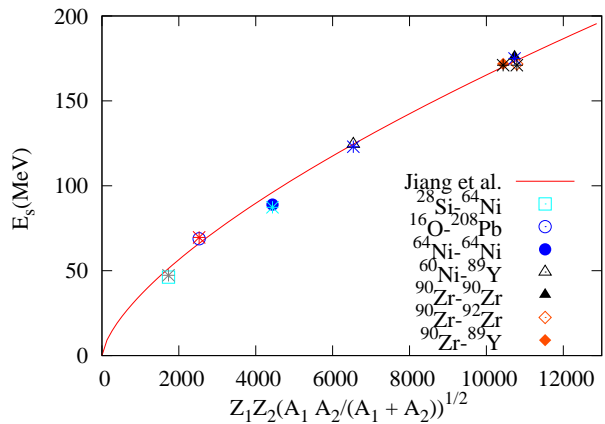}
\caption{(Color online) The threshold energy $E_s$ for deep subbarrier hindrance 
for several systems, determined with the two slope fit to the 
experimental fusion cross sections, as a function of the 
parameter $Z_1Z_2\sqrt{A_1A_2/ (A_1+A_2)}$. 
The solid curve is the empirical function given by 
Eq. (\ref{eq:E_s}), while the stars denote the 
``experimental'' values defined as the maximum energy of the astrophysical 
S-factors\cite{bt9}. }
\label{fig:2}
\end{center}
\end{figure}

\begin{table*}[hbt]
\caption{The threshold energy $E_s$ for deep subbarrier fusion hindrance for 
several systems, obtained with the two slope fit to the experimental 
fusion cross sections. $a_>$ and $a_<$ are the diffuseness parameters 
in the Woods-Saxon potential used to fit the subbarrier and the deep 
subbarrier regions of fusion cross sections. 
$\zeta$ is defined as $\zeta= Z_1Z_2\sqrt{A_1A_2/ (A_1+A_2)}$, 
in which $Z_i$ and $A_i~(i=1,2)$ are the charge and the mass numbers of 
the nucleus $i$. 
$E_s^{\rm (exp)}$ and $E_s^{\rm (emp)}$ are the ``experimental'' threshold 
energy \cite{bt9} and the 
empirical energies given by Eq. (\ref{eq:E_s}), respectively. 
$V_{\rm KNS}$ is the potential energy at the touching configuration 
\cite{bt1} estimated 
with the KNS potential. 
All the energies are shown in units of MeV, while the lengths are 
in units of fm. }
\begin{center}
\label{table}
\begin{tabular}{c|c|ccc|ccc}
\hline
\hline
systems & $\zeta$ & 
$a_>$ & $a_<$ & $E_s$ & 
$E^{\rm (exp)}_s$ &  $E^{\rm (emp)}_s$ & $V_{\rm KNS}$ \\ 
\hline
$^{28}$Si+$^{64}$Ni & 1730.05 & 0.71 & 0.99 & 46.2 & 
47.3$\pm$0.9 & 51.3 & 43.9\\ 
$^{16}$O+$^{208}$Pb & 2528.55 & 0.87 & 0.94 & 68.8 & 69.6 & 66.1 & 70.5\\ 
$^{64}$Ni+$^{64}$Ni & 4434.97 & 0.76 & 0.9 & 88.92 & 
87.3$\pm$0.9 & 96.1 & 89.0\\ 
$^{60}$Ni+$^{89}$Y & 6537.33 & 0.74 & 0.815 & 124.5 & 123$\pm$1.2 & 124.5 & 
125.4\\ 
$^{90}$Zr+$^{89}$Y & 10435.5 &  0.76 & 0.87 & 171.8 & 171$\pm$1.7 & 170.3 &
175.2\\ 
$^{90}$Zr+$^{90}$Zr & 10733.1 &  0.56 & 0.76 & 176.1 & 
175$\pm$1.8 & 173.2 & 179.9\\ 
$^{90}$Zr+$^{92}$Zr & 10791.9 &  0.53 & 0.78 & 171.7 & 
171$\pm$1.7 & 173.9 & 179.1 \\
\hline
\hline
\end{tabular}
\end{center}
\end{table*}

Figure \ref{fig:2} shows the threshold energies thus obtained as a function 
of $Z_1Z_2\sqrt{A_1A_2/(A_1+A_1)}$. The figure also shows the 
threshold energy for 
$^{28}$Si+$^{64}$Ni\cite{bt4}, $^{64}$Ni+$^{64}$Ni\cite{bt8}
$^{16}$O+$^{208}$Pb\cite{ANU07}, 
$^{60}$Ni+$^{89}$Y\cite{bt8}, $^{90}$Zr+$^{90}$Zr\cite{bt6}, 
$^{90}$Zr+$^{92}$Zr\cite{bt6}, 
and $^{90}$Zr+$^{89}$Y \cite{bt6}systems. 
For comparison, the figure also shows 
the empirical systematics given by Eq. (\ref{eq:E_s}) with the solid line 
and the ``experimental'' data defined as the maximum energy of the S-factor 
\cite{bt9} by the crosses. 
These values are summarized in Table \ref{table}, together with the 
potential energy at the touching point \cite{bt1} 
estimated with the KNS potential. 
One can see that 
the values of the threshold energy defined in our way 
are in 
good agreement with those defined as the maximum of the 
astrophysical S-factor as well as with the potential energy at the 
touching configuration. 

Let us next discuss briefly the asymptotic 
energy shift for deep subbarrier fusion reactions. 
This quantity was introduced by Aguiar {\it et al.} \cite{bt3} as 
a measure of subbarrier enhancement of fusion cross sections. 
It was defined as 
an extra energy needed to fit the experimental 
fusion cross sections with respect to a single-channel potential model calculation. 
It has been argued that 
the calculated 
fusion cross sections 
have approximately the same exponential energy dependence as the experimental 
data in the subbarrier energy region, 
but are shifted in energy by a constant amount \cite{bt3}. 
In connection to the deep subbarrier fusion hindrance, it 
may be interesting to revisit this representation. 

\begin{figure}[tbhp]
\begin{center}
\includegraphics[width=1\linewidth, clip]{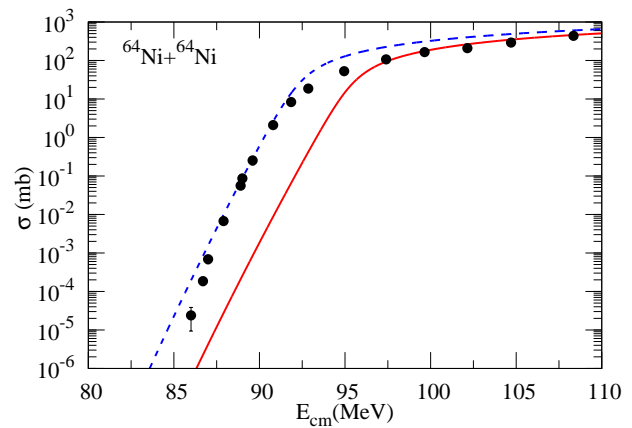}
\caption{(Color online) 
Fusion excitation functions for the 
$^{64}$Ni+$^{64}$Ni system. 
The dashed and the solid lines are results of single-channel 
potential model calculations which fit the experimental data in the subbarrier region and 
at energies above the Coulomb barrier, respectively. 
The experimental data are taken from 
Refs. \cite{bt8}. }
\label{fig:3}
\end{center}
\end{figure} 

\begin{figure}[tbhp]
\begin{center}\leavevmode
\includegraphics[width=1\linewidth, clip]{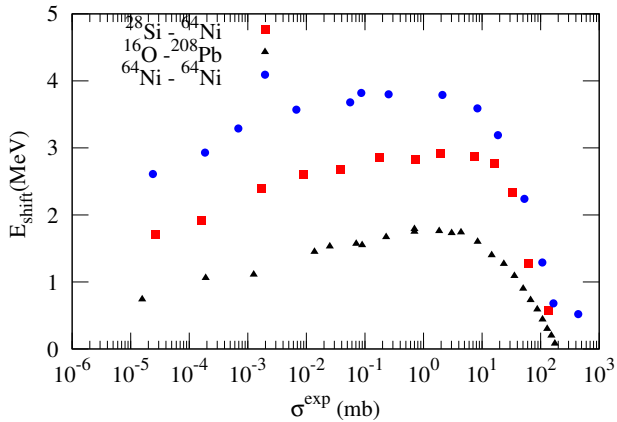}
\caption{(Color online) 
The asymptotic energy shift as a function of fusion cross section for 
$^{28}$Si+$^{64}$Ni (the filled squares), 
$^{16}$O+$^{208}$Pb (the filled triangles), 
and 
$^{64}$Ni+$^{64}$Ni (the filled circles) systems.}
\label{fig:4}
\end{center}
\end{figure}

In order to define the asymptotic energy shift, we first adjust the value of 
$V_0$ and $R_0$ in the Woods-Saxon potential, keeping the same value for the 
diffuseness parameter $a$ as the one which has been obtained to fit to the subbarrier 
fusion cross sections (see $a_>$ in Table I), so that the experimental fusion cross 
sections at high energies, that is, those above $\sigma > 100$ mb, 
can be approximately reproduced (see Fig. \ref{fig:3}). 
We then define the asymptotic energy shift as a difference between the solid line in 
Fig. \ref{fig:3} and the experimental data for a fixed value of fusion cross section. 
Figure \ref{fig:4} shows the asymptotic energy shift so extracted for several systems 
as a function of corresponding fusion cross section. 
As one can see, 
the asymptotic energy shift is nearly 
constant in the range of 0.1 mb $\alt \sigma \alt$ 1 mb, 
in accordance to the previous conclusion 
by Aguiar {\it et al.}\cite{bt3}. 
However, in the deep subbarrier region, the asymptotic energy shift start decreasing 
as the fusion cross sections decrease, reflecting the fact that the fusion cross sections have a 
different exponential slope from that in the subbarrier region, as shown in 
Fig. \ref{fig:1}. 

In summary, we have studied the energy dependence of heavy-ion fusion cross sections 
at deep subbarrier energies using the recent experimental data. 
To this end, we employed a one-dimensional potential model. 
We have shown that the asymptotic energy shift 
is almost a constant in the subbarrier region, but 
it decreases with decreasing energies in the deep subbarrier region. 
This is a clear manifestation of the hindrance phenomenon of deep sub-barrier fusion.
In order to see at which energy the deep subbarrier hindrance takes place, we estimated 
the threshold energy with a two-slope fit procedure. 
That is, we defined the threshold energy as an intersect of two fusion excitation functions, 
which fit the experimental fusion cross sections either in the subbarrier energy region or in 
the deep subbarrier energy region. 
We have shown that the threshold energies so defined are in good agreement 
with those estimated from the maximum of astrophysical $S$-factor. 

The definition for the threshold energy proposed in this paper is 
complementary to the one with the maximum of astrophysical S-factor. 
As we have shown in this paper, both the definitions provide a similar value of threshold energy 
as the potential energies at the touching configuration. 
This strongly suggests that the dynamics after the touching plays an important role in deep subbarrier fusion 
reactions, changing the exponential slope of fusion cross sections and at the same time 
making the astrophysical S-factor take the maximum, although 
it is an open question why and how 
the dynamics after the touching leads to the maximum of astrophysical S-factor.

\bigskip

We thank T. Ichikawa for useful discussions. 
This work was supported by the Japanese
Ministry of Education, Culture, Sports, Science and Technology
by Grant-in-Aid for Scientific Research under
the program number  (C) 22540262.

\end{document}